\documentstyle[12pt]{article}

\title{Study of a toy model and its relation to the
Hubbard model with infinite range hopping}

\author{Flavio S. Nogueira \thanks{e-mail: flaviono@cbpfsu1.cat.cbpf.br}
\\
Centro Brasileiro de Pesquisas Fisicas - CBPF \\
Rua Dr. Xavier Sigaud, 150, Rio de Janeiro, RJ 22290-180 \\
BRAZIL \\
Enrique V. Anda \\
Universidade Federal Fluminense - UFF \\
Instituto de Fisica (Campus da Praia Vermelha) \\
Av. Litor\^anea s/n - Gragoat\'a \\
24210-340 Niter\'oi - RJ - BRAZIL}

\begin{document}

\maketitle

\begin{abstract}
A toy model of strongly correlated fermions is studied
using Green function and functional integration methods.
The model exhibits a metal-insulator transition as
the interaction is varied.
In the case of unrestricted hopping is established an
equivalence of the model with the Hubbard model with
infinite range hopping. The generalization to the case with
$N$ components is made.
\end{abstract}

\section{Introduction} \setcounter{equation}{0}

The simplest model to describe strong electron correlations is the
Hubbard model \cite{Hub}. Unfortunately, the
only exact solution available corresponds to the one dimensional case
\cite{LW}. Thus, to understand its physical properties we have to relay
upon
approximate solutions which in many instances are highly non-trivial. Another
alternative is
to formulate exactly soluble toy models which share some properties with
realistic many body models. The importance of the toy models lies in
the fact that many interesting properties of complicated strongly
correlated electron systems are simulated in a compreensive way.

A simple toy model was proposed by Hatsugai and
Kohmoto \cite{HK} (to be referred from now on as the 'HK model'). This model
is exactly solvable in a very simple way and have the same atomic and band
limits as the Hubbard model. Also, as Hatsugai and Kohmotto have shown, the
model describes a metal-insulator transition
(MIT). This MIT scenario of the HK model was studied from a
scaling point of view by Continentino \cite{Cont} who also formulated a boson
version of the HK model. A closely related toy model was also discussed by
Baskaran \cite{Bas}.
There is also an atractive version of the HK model
which was proposed recently by Mattis and Bedersky \cite{MB}.

Another different toy model
consisting by a Hubbard model with an infinte range
hopping has been solved exactly \cite{vDV}.
This model has the
same atomic limit as the Hubbard model but does not have obviously the same
band limit since hopping is unrestricted. As shown in ref. \cite{vDV},
this model is an insulator at half-filling for any  on site
Coulombian repulsion $U>0$.

In this work we discuss the HK model using Green function and
integral functional methods. It is
shown that the HK model and the Hubbard model with infinite range hopping
are equivalent in the thermodynamic limit
if the hopping in the HK model is assumed to be unrestricted.
We also solve exactly the $N$ component HK model
for any {\it finite} $N$. This
solution is obtained by an {\it exact} evaluation of the functional integral
representation of the model. From this solution one obtains also the exact
solution of the $N$ component Hubbard model with infinite range hopping.
For the sake of completeness, we also dicuss a lattice fermion model
where the interaction, rather than the hopping is of infinite range.

\section{The HK model} \setcounter{equation}{0}

The HK model is described by the Hamiltonian:

\begin{equation}
H=\sum_{\vec{k}}\sum_{\sigma}[\epsilon(\vec{k})-\mu]n_{\vec{k}\sigma}
+U\sum_{\vec{k}}n_{\vec{k}\uparrow}n_{\vec{k}\downarrow}
\end{equation}
where $n_{\vec{k}\sigma}{\equiv}c_{\vec{k}\sigma}^{\dag}c_{\vec{k}\sigma}$
and $\epsilon(\vec{k})=-2t\sum_{i=1}^{d}{\cos}k_{i}$. $\mu$ is
the chemical potencial. The dispersion
$\epsilon(\vec{k})$ corresponds to a nearest neighbour hopping on a
hypercubic lattice in $d$ dimensions. Note that, in contrast to the
Hubbard model, the Coulombian repulsion is local in $\vec{k}$-space rather
than in real space. Since the Hamiltonian is local in $\vec{k}$-space and
the interacting part commutes with the non-interacting one, it is
straightforward to solve it. For instance, the exact free energy density is

\begin{equation}
\label{f}
f=-\frac{1}{L\beta}\sum_{\vec{k}}\ln\{1+2e^{\beta[\mu-\epsilon(\vec{k})]}
+e^{\beta[2\mu-2\epsilon(\vec{k})-U]}\}
\end{equation}
where $\beta=1/T$, $T$ being the temperature, and $L$ is the number of lattice
sites. Note that in the zero bandwidth limit we
obtain the free energy density of
the atomic limit of the Hubbard model. Also, in the limit $U=0$ we get the
band limit.

The Matsubara Green function is defined by
$G_{\sigma}(\vec{k},\tau)=-<Tc_{\vec{k}\sigma}(\tau)c_{\vec{k}\sigma}^{\dag}(0)>$.
In frequency representation it is given straightforwardly by

\begin{equation}
\label{G}
G_{\sigma}(\vec{k},\omega_{n})=\frac{1-n_{\vec{k}\overline{\sigma}}}
{i\omega_{n}
+\mu-\epsilon(\vec{k})}+\frac{n_{\vec{k}\overline{\sigma}}}{
i\omega_{n}+\mu
-\epsilon(\vec{k})-U},
\end{equation}
where $\omega_{n}=(2n+1){\pi}T$, $n{\in}Z$, and

\begin{equation}
\label{n}
n_{\vec{k}\overline{\sigma}}=\frac{e^{\beta[\mu-
\epsilon(\vec{k})]}+e^{\beta[2\mu
-2\epsilon(\vec{k})-U]}}{1+2e^{\beta[\mu-\epsilon(\vec{k})]}
+e^{\beta[2\mu-2\epsilon(\vec{k})-U]}}.
\end{equation}
Note that we are assuming a paramagnetic phase, that is,
$n_{\vec{k}\uparrow}=n_{\vec{k}\downarrow}$. At $T=0$ Eq.(\ref{n}) assumes the
form of a step function:

\begin{equation}
n_{\vec{k}\sigma}(T=0)=\theta(\mu_{0}-\epsilon(\vec{k}))[\frac{1}{2}
\theta(U-|\epsilon(\vec{k})-\mu_{0}|)+\theta(|\epsilon(\vec{k})-\mu_{0}|-U)],
\end{equation}
where $\mu_{0}$ is the chemical potential at $T=0$. $\theta(x)$ is the usual
Heaviside function. This occupation number  has two
discontinuous jumps. An immediate consequence of this is that
Luttinger's theorem does not hold.
Note that the exact Green function for the HK model share the same atomic
and band limits with the Hubbard model.

As in the case of the Hubbard model, a half-filling condition is obtained by
setting $\mu=U/2$. At half-filling and $T=0$ we have that the Green function
becomes

\begin{equation}
\label{g}
G_{\sigma}(\vec{k},\omega)=\frac{i\omega}{[i\omega-\epsilon(\vec{k})]^{2}
-\frac{U^{2}}{4}}.
\end{equation}
In above, $\omega$ is not more a Matsubara frequency but an Euclidian
frequency, that is, we still have an analytic continuation to
imaginary time but in zero temperature. Thus, the frequency $\omega$ is
continuous and not discrete.

{}From the poles of Eq.(\ref{g}) we obtain two bands in straight analogy with
the case of the Hubbard model where we have the so called lower
Hubbard band and
the upper Hubbard band. The poles gives the energy bands:

\begin{eqnarray}
E_{-} & = & \epsilon(\vec{k})-\frac{U}{2} \\
E_{+} & = & \epsilon(\vec{k})+\frac{U}{2}
\end{eqnarray}
Note that the gap has a size which depends on $U$. If the two bands are
separated by a nonzero gap we have an insulator. The value of U which gives
a zero  gap is determined by demanding that the top of the
lower band coincides with the bottom of the lower one. Thus we find that
the critical value of $U$ that signals a MIT is given by $U_{c}=4td=W$.
Thus for $U<U_{c}$ the system is metallic. Note that in the
HK model we have a MIT at any dimension while in the case of the Hubbard model
it is well known that in one dimension no MIT exists at half-filling \cite{LW}.

It is also interesting to study the behavior
of the spin suscetibility. This is given by the response
function:

\begin{equation}
\chi(\vec{k},\nu_{n})=-\mu_{B}^{2}\Pi(\vec{k},\nu_{n})
\end{equation}
where $\mu_{B}$ is the Bohr magneton and $\Pi(\vec{k},\nu_{n})$ is the
polarization Green function which is given by:

\begin{equation}
\Pi(\vec{k},\nu_{n})=\frac{1}{L\beta}\sum_{\vec{q}}\sum_{\omega_{n}}
G(\vec{k}+\vec{q},\nu_{n}+\omega_{n})G(\vec{q},\omega_{n}),
\end{equation}
where $\nu_{n}=2n{\pi}T$, $n{\in}Z$, is a Bose Matsubara frequency. We are
omitting spin indices for simplicity. The Matsubara sum is over a fermion
frequency. We will use the Green functions at half-filling.
We evaluate the suscetibility at zero temperature in the static limit and at
the nesting wave vector $\vec{Q}=(\pi,...,\pi)$. Performing the Matsubara
sum and taking the zero temperature limit gives

\begin{eqnarray}
\label{chi}
\chi(\vec{Q},0) & = & -\frac{\mu_{B}^{2}}{4}\int_{1BZ}\frac{d^{d}q}{(2\pi)^{d}}
\left(\frac{\theta(U/2+\epsilon(\vec{q}))-\theta(U/2-\epsilon(\vec{q}))}
{2\epsilon(\vec{q})}\right. \nonumber \\
 &   &
\mbox{}+\frac{\theta(U/2+\epsilon(\vec{q}))-\theta(-\epsilon(\vec{q})-U/2)}
{2\epsilon(\vec{q})+U}+\frac{\theta(\epsilon(\vec{q})-U/2)-\theta(U/2
-\epsilon(\vec{q}))}{2\epsilon(\vec{q})-U} \nonumber \\
 &   &
\mbox{}+\left.\frac{\theta(\epsilon(\vec{q})-U/2)-\theta(-\epsilon(\vec{q})-U/2)}
{2\epsilon(\vec{q})}\right)
\end{eqnarray}
When $U=U_{c}=W$ the integral of the second term in Eq.(\ref{chi}) is
divergent for $\vec{q}=0$ ("infrared divergence") while the integral of the
third term is divergent for $\vec{q}=\vec{Q}$ ("ultraviolet divergence").

Let us evaluate explicitly the spin susceptibility. If we use a square
density of states, $\rho(\omega)=(1/W)\theta(W/2-\omega)\theta(\omega+W/2)$,
 and consider $U>U_{c}$ we get

\begin{equation}
\chi(\vec{Q},0)=-\frac{\mu_{B}^{2}}{4W}\ln\left(\frac{U+U_{c}}{U-U_{c}}\right)
\end{equation}
Thus we have in fact that the susceptibility
diverges as $U$ approaches $U_{c}$.
Note that we do not have a simple power law behavior for the suscetibility.
This should be contrasted with other descriptions of the MIT. For exemple,
the Brinkman and Rice transition \cite{BR} gives a power law  behavior for the
static suscetibility, which is calculated for a zero value of the
reciprocal lattice vector rather than in the nesting wave vector.

The analysis of this section allows us to conclude that the MIT in the HK
model is a second order transition with mass gap
$\Delta=U-U_{c}$.
Note that $\Delta$ plays a role similar to the $r_{0}$
coupling associated to the quadratic part in the Ginsburg-Landau Hamiltonian.
Remember that in a Ginsburg-Landau theory it is assumed that
$r_{0}{\sim}T-T_{c}$ where $T_{c}$ is the critical temperature associated to
the second order transition. Here we have $U$ rather than $T$.

\section{The Hubbard model with infinite range hopping}
\setcounter{equation}{0}

The kinetic energy of the HK model is identical to the one of the
Hubbard model. Thus it is natural to ask about the differences of behavior
between the HK model and the Hubbard model as we manipulate or change the
hopping matrix elements. In this section we will consider the HK model with
infinite range hopping (IRH). The IRH Hubbard model was
already studied and solved exactly
\cite{vDV}. We will solve exactly the IRH HK model
and show that the solution is the same in the thermodynamic limit, to
that of the IRH Hubbard model. This establishes the equivalence
between the IRH HK model and the IRH Hubbard model. In the case of
infinite range hopping it does not matter if the electrons interact
locally in $\vec{k}$-space or in real space.
In both situations the solution
is the same.

When the hopping is of infinite range the dispersion is given
by $\epsilon(\vec{k})=-Lt\delta_{\vec{k},0}$ \cite{vDV}. The exact free
energy density for the IRH HK model is obtained by substituting this
dispersion in the exact expression Eq.(\ref{f}) and taking the
thermodynamic limit. It is readily obtained that

\begin{equation}
\label{f1}
f_{IRH}=-2t-\frac{1}{\beta}\ln[1+2e^{\beta\mu}+e^{\beta(2\mu-U)}],
\end{equation}
which is the same expression that corresponds to
the free energy density of the
IRH Hubbard model \cite{vDV}.

Also, by substituting the IRH dispersion in Eq.(\ref{G}) and taking
the thermodynamic limit one obtains the
exact Green function for the IRH HK model:

\begin{equation}
\label{GIRH}
G_{IRH}(\vec{k},\omega_{n})=\frac{(1-\delta_{\vec{k},0})}{Z_{0}}\left(
\frac{1+e^{\beta\mu}}{i\omega_{n}+\mu}+\frac{e^{\beta\mu}+e^{\beta(2\mu-U)}}{
i\omega_{n}+\mu-U}\right),
\end{equation}
$Z_{0}$ being the atomic partition function per site.
Since the free energy of the IRH HK model coincides with that of the IRH
Hubbard model, the above expression gives the exact Green function of
the IRH Hubbard model.
It turns out that
the two models have the same perturbative expansion in the
thermodynamic limit.
Note that the $\vec{k}=0$ mode
does not propagate while all the others $\vec{k}$ modes have
$\vec{k}$ independent propagators. Note also that the $\vec{k}{\neq}0$
modes have their propagators identical to the atomic limit Green
function. This behavior is manifested also in the expression for the
free energy, Eq.(\ref{f1}), where we have a term given by the
atomic solution corresponding to the modes $\vec{k}{\neq}0$ and a term
$-2t$ corresponding to the $\vec{k}=0$ mode. This means that in the
IRH regime the $\vec{k}=0$ and $\vec{k}{\neq}0$ modes separate.

{}From Eq.(\ref{GIRH}) one obtains the electron density:

\begin{eqnarray}
\label{nIRH}
n & = &
\frac{2}{L\beta}\sum_{\vec{k}}\sum_{n}e^{-i\omega_{n}0+}G_{IRH}(\vec{k},
\omega_{n}) \nonumber \\
  & = & \mbox{}\frac{2[e^{\beta\mu}+e^{\beta(2\mu-U)}]}{1+2e^{\beta\mu}
+e^{\beta(2\mu-U)}}
\end{eqnarray}
The above relation can also be obtained directly from the expression for
the free energy as $n=-({\partial}f_{IRH}/\partial\mu)_{\beta}$.
Solving Eq.(\ref{nIRH}) for $\mu$ one obtains

\begin{equation}
\mu=U+\beta^{-1}\ln\left[
\frac{\sqrt{(1-n)^{2}+n(2-n)e^{-{\beta}U}}-(1-n)}{2-n}\right]
\end{equation}
in agreement with reference \cite{vDV}.

{}From Eq.(\ref{GIRH}) we obtain the density of doubly occupied sites
$\overline{d}$:

\begin{equation}
\overline{d}(\beta,U,\mu)=\frac{e^{\beta(2\mu-U)}}{1+2e^{\beta\mu}+
e^{\beta(2\mu-U)}}
\end{equation}
The above expression in function of $\mu$ is far more compact than that one
obtained in ref.\cite{vDV} which has been written as a function of $n$. When
$\mu=U/2$, which corresponds to $n=1$, one has

\begin{equation}
\overline{d}=\frac{1}{2(e^{\frac{{\beta}U}{2}}+1)},
\end{equation}
which again agrees with ref.\cite{vDV}.

\section{The HK model with N components} \setcounter{equation}{0}

In section 2 we solved exactly the HK model with $N=2s+1=2$ components ($s$
is the spin). In this section we show that this model is also
exactly soluble for an arbitrary number of components.

Let us define the $N$ component spinors:

\begin{equation}
\psi_{\vec{k}}\equiv\left(
\begin{array}{c}
\psi_{\vec{k}1} \\
\cdot \\
\cdot \\
\cdot \\
\psi_{\vec{k}N}
\end{array}
\right),
\end{equation}

\begin{equation}
\overline\psi_{\vec{k}}\equiv\left(
\begin{array}{ccc}
\psi_{\vec{k}1}^{*} & ... & \psi_{\vec{k}N}^{*}
\end{array}
\right).
\end{equation}
Each component of the expression above
is a Grassmann field. We will evaluate the exact
partition function by writing it as an functional integral over the fields
defined above. The partition function is given by

\begin{equation}
Z=\int\prod_{\vec{k}}D\overline\psi_{\vec{k}}D\psi_{\vec{k}}e^{-S[\overline
\psi_{\vec{k}},\psi_{\vec{k}}]}
\end{equation}
where the action $S$ is

\begin{equation}
S[\overline\psi_{\vec{k}},\psi_{\vec{k}}]=\int_{0}^{\beta}d\tau
\sum_{\vec{k}}[\overline\psi_{\vec{k}}(\partial_{\tau}-\mu-U/2+
\epsilon(\vec{k}))\psi_{\vec{k}}+\frac{U}{2}(\overline\psi_{\vec{k}}
\psi_{\vec{k}})^{2}],
\end{equation}
where it is understood that the fields are time dependent (Matsubara time) and
that they satisfy antiperiodic boundary conditions in Matsubara time.

We eliminate the quartic term by means of a Hubbard-Stratonovich
transformation,
getting the new action,

\begin{equation}
S'=\int_{0}^{\beta}d\tau\sum_{\vec{k}}[\overline\psi_{\vec{k}}(\partial_{\tau}
-\mu-U/2+\epsilon(\vec{k})-iU\phi_{\vec{k}})\psi_{\vec{k}}+(U/2)\phi_{\vec{k}}
^{2}]
\end{equation}
where $\phi_{\vec{k}}$ is an auxiliary Bose field. Since the new action is
quadratic in the Fermi fields, it is straightforward integrate out these and
obtain the following effective action:

\begin{equation}
S_{eff}=-N\ln\det[\partial_{\tau}-\mu-U/2+\epsilon(\vec{k})-iU\phi_{\vec{k}}]+
\frac{U}{2}\int_{0}^{\beta}d\tau\sum_{\vec{k}}\phi_{\vec{k}}^{2}
\end{equation}
We can evaluate exactly the determinant appearing in the above equation by
solving the differential equation:

\begin{equation}
\label{ode}
[\partial_{\tau}-\mu-U/2+\epsilon(\vec{k})-iU\phi_{\vec{k}}(\tau)]f_{n}(\vec{k},
\tau)=\alpha_{n}(\vec{k})f_{n}(\vec{k},\tau)
\end{equation}
with $f_{n}$ satisfying the antiperiodic boundary condition
$f_{n}(\vec{k},0)=-f_{n}(\vec{k},\beta)$. The Eq.(\ref{ode}) is easy to solve
and the solution is given by

\begin{equation}
f_{n}(\vec{k},\tau)=c\exp\left(\int_{0}^{\tau}d{\tau}'[\mu+U/2-\epsilon(\vec{k})+
iU\phi_{\vec{k}}({\tau}')+\alpha_{n}(\vec{k})]\right),
\end{equation}
$c$ being an arbitrary constant.
By applying the antiperiodic boundary condition to this solution we get

\begin{equation}
\alpha_{n}(\vec{k})=-i\omega_{n}-\mu-U/2+\epsilon(\vec{k})-i(U/\beta)
\int_{0}^{\beta}d\tau\phi_{\vec{k}}
\end{equation}
where $\omega_{n}$ is a Fermi Matsubara frequency. Thus, the determinant
which appears in the effective action is given by the product of the
$\alpha_{n}$'s. Therefore,

\begin{equation}
\label{Seff}
S_{eff}=-N\sum_{n=-\infty}^{+\infty}\sum_{\vec{k}}\ln[-i\omega_{n}-\mu-U/2+
\epsilon(\vec{k})-i(U/\beta)\int_{0}^{\beta}d\tau\phi_{\vec{k}}(\tau)]
+\frac{U}{2}\int_{0}^{\beta}d\tau\phi_{\vec{k}}^{2}(\tau)
\end{equation}
The Matsubara sum in Eq.(\ref{Seff}) is easily done by standard methods. The
partition function is then written in the form:

\begin{equation}
Z=\int\prod_{\vec{k}}D\phi_{\vec{k}}\left\{1+e^{\beta[\mu+U/2-\epsilon(\vec
{k})]}e^{iU\int_{0}^{\beta}d\tau\phi_{\vec{k}}(\tau)}\right\}^{N}
e^{-\frac{U}{2}\int_{0}^{\beta}d\tau\phi_{\vec{k}}^{2}(\tau)}
\end{equation}
By expanding the term between braces and performing the Gaussian integrals
one obtains the {\it exact} expression for the partition function of the
$N$ component HK model:

\begin{equation}
\label{ZN}
Z=\prod_{\vec{k}}\sum_{n=0}^{N}\frac{N!}{n!(N-n)!}e^{n\beta[\mu-\epsilon(
\vec{k})]}e^{\beta\frac{n(n-1)U}{2}}
\end{equation}
{}From Eq.(\ref{ZN}) one obtains the free energy density:

\begin{equation}
\label{fN}
f=-\frac{1}{L\beta}\sum_{\vec{k}}\ln\left\{\sum_{n=0}^{N}\frac{N!}{
n!(N-n)!}e^{n\beta[\mu-\epsilon(\vec{k})]}e^{\beta\frac{n(n-1)U}{2}}\right\}.
\end{equation}
Note that in the case of $N=2$ the result given in Eq.(2.2) is recovered.
Also, in the zero bandwidth limit one obtains the atomic limit of a
Hubbard model with $N$ components.

It is worth to emphasize that the above result is {\it exact} for any
finite $N$. We can also look for the solution at large $N$. The limit
$N\rightarrow\infty$ of Eq.(\ref{fN}) is not obvious because of the
factorials. However, we can obtain the large $N$ solution by a saddle point
evaluation of the functional integral with the effective action (\ref{Seff}).
In order to perform this saddle point evaluation it is useful to make the
rescaling $U\rightarrow\frac{U}{N}$,
$\phi_{\vec{k}}{\rightarrow}N\phi_{\vec{k}}$.
In this way we get that $S_{eff}=N\overline{S_{eff}}$ where
$\overline{S_{eff}}$ is $N$ independent (note that the term $U/2$ inside the
logarithm is rescaled to $U/(2N)$ which is zero at large $N$). In this case
the saddle point solution corresponds to the exact solution when
$N\rightarrow\infty$. The large $N$ solution is found to be just the
Hartree-Fock approximation for the HK model.

{}From the exact expression (\ref{fN}) it is possible to obtain the exact free
energy density for the $N$ component IRH Hubbard model. It is given by

\begin{equation}
f_{IRH}(N)=-Nt-\frac{1}{\beta}\ln\left[\sum_{n=0}^{N}\frac{N!}{n!(N-n)!}
e^{n\beta\mu}e^{\beta\frac{n(n-1)U}{2}}\right].
\end{equation}

\section{The infinite range interaction case} \setcounter{equation}{0}

Let us write the interaction of the HK model in the real space
representation. This is achieved by Fourier transformation:

\begin{equation}
c_{\vec{k}\sigma}=\frac{1}{\sqrt{L}}\sum_{i}\exp{(-i\vec{k}\cdot
\vec{R}_{i})}c_{i\sigma}.
\end{equation}
It is readily obtained that the interaction part is given by

\begin{equation}
\label{5}
H_{1}=\frac{U}{L}\sum_{ijkl}\delta_{i+k,j+l}c^{\dag}_{i\uparrow}
c_{j\uparrow}c^{\dag}_{k\downarrow}c_{l\downarrow}.
\end{equation}
The above interaction is obviously of infinite range. We note that
it is not necessary to consider only interaction between opposite
spins. Generalizing the above interaction in a such a
way as to include interaction between like spins correspond to
a shift in the chemical potential. In this situation we have a
particular case of the HK model which consist of a model with
Hamiltonian given by

\begin{equation}
\label{5.1}
H=-t\sum_{<i,j>}\sum_{\sigma}c_{i\sigma}^{\dag}c_{i\sigma}+h.c.
-\mu\sum_{i}\sum_{\sigma}n_{i\sigma}+\frac{U}{2L}\sum_{i,j}n_{i}n_{j},
\end{equation}
where $n_{i}=\sum_{\sigma}n_{i\sigma}$.
Since the interaction is the
same for every pair of sites in the lattice we can write
$\sum_{i,j}n_{i}n_{j}=(\sum_{i}n_{i})^{2}$. It is straightforward to
write the above Hamiltonian in $\vec{k}$-space as

\begin{equation}
\label{5.2}
H=\sum_{\vec{k}}\sum_{\sigma}[\epsilon(\vec{k})-\mu]n_{\vec{k}\sigma}
+\frac{U}{2L}(\sum_{\vec{k}}n_{\vec{k}})^{2}.
\end{equation}

We write the partition function in the same way as in the last section,
that is, by going to an integral functional representation, the
fermionic fields being substituted by Grassmann fields. As before, the
quartic term in the action is eliminated by applying a
Hubbard-Stratonovich transformation. We obtain the following action
involving an auxiliary Bose field $\phi$:

\begin{equation}
\label{5.3}
S=\int_{0}^{\beta}d\tau\{\sum_{\vec{k}}\sum_{\sigma}\overline{\psi}_{
\vec{k}\sigma}[\partial_{\tau}-\mu+\epsilon(\vec{k})-iU\phi]\psi_{\vec{k}
\sigma}+\frac{UL}{2}\phi^{2}\}.
\end{equation}
By integrating the fermions and carrying the calculations
in a way similar to  the one in the preceeding section, we get an
effective action $S_{eff}=L\overline{S}_{eff}$, where
$\overline{S}_{eff}$ is given by

\begin{equation}
\label{5.4}
\overline{S}_{eff}=-2\int_{-\infty}^{\infty}d\epsilon\rho_{0}(\epsilon)
\sum_{n}\ln[-i\omega_{n}-\mu+\epsilon-i\frac{U}{\beta}\int_{0}^{\beta}
d\tau\phi(\tau)]+\frac{U}{2}\int_{0}^{\beta}d\tau\phi^{2}(\tau),
\end{equation}
where $\rho_{0}(\epsilon)$ is the bare density of states. Therefore, we
write the partition function in the following form:

\begin{equation}
\label{5.5}
Z=N{\int}D{\phi}e^{-L\overline{S}_{eff}[\phi]},
\end{equation}
where $N$ is a normalization factor.
{}From Eq.(\ref{5.5}) we see that in the thermodynamic limit the saddle point
of the action corresponds to the exact solution of the problem. The
saddle point is given by an imaginary number in the repulsive case:
$\phi_{0}=in_{0}$, $n_{0}$ being a real number in the interval
$[0,2]$. In the attractive case the saddle point is a real number because
the resulting Hubbard-Stratonovich transformation will not involve
the imaginary unity $i$ multiplying the linear term in $\phi$ as in the
preceding equations. The saddle point equation is given by

\begin{equation}
\label{5.6}
n_{0}=\frac{2}{\beta}\sum_{n}\int_{-\infty}^{\infty}d\epsilon
\frac{\rho_{0}(\epsilon)}{i\omega_{n}+\mu-\epsilon-Un_{0}}.
\end{equation}
Eq.(\ref{5.6}) is just a Hartree-Fock aproximation. This means that in the
case of an infinite range interaction of the form described
in Eq.(\ref{5.1}) the Hartree-Fock approximation
is the exact solution of the problem. Note that in
contrast to the case of the
HK model, we have a Fermi liquid in the present situation.

\section{Conclusion}

Strongly correlated electron systems are very important in today
condensed matter physics. Unfortunately, even the simplest models
developed in this field are very complicated to handle. For this
reason, many important features cannot be understood exactly. For
example, the MIT in the Hubbard model can be studied only
approximately. The exactly soluble one dimensional Hubbard model
does not exibits a MIT at half-filling. However, the MIT in the Hubbard
model can be simulated through the study of exactly solvable toy
models like the one we have discussed in this paper, the HK model.
We discussed the HK model within the framework of the Green function
and functional integration formalism. The metal-insulator transition
at half-filling is easily characterized from the structure of the poles of the
Green function. The exact Green function is used for evaluate the
static susceptibility at the nesting wave vector at zero
temperature. It was found that the susceptibility diverges as $U$
approaches the critical value $U_{c}$, signaling a localization
transition. This also happens in the Brinkman-Rice approximation in
the Hubbard model. However, in the Brinkman-Rice case the
static susceptibility is evaluated at zero wave vector rather than in the
nesting vector. Also, in the case treated here we do not have a power law
behavior of the susceptibility.

Another interesting result is that the HK model with infinite range hopping is
equivalent to the
Hubbard model with infinite range hopping, giving the same
physical information. For example, the free energy of both models
coincide. Actually, this is an expected result once the IRH Hubbard model
separate the $\vec{k}=0$ modes of the $\vec{k}{\neq}0$ modes and the HK
model merges the exact atomic and band limits of the Hubbard model.
Thus, we have shown that the HK model shares one more propertie with the
Hubbard model. In the limit of unrestricted hopping the two models are
equivalent.

Finally, the HK model admits a further generalization. It can be solved
exactly even in the case of an arbitrary number of components.

\end{document}